\documentclass[final,1p,times]{elsarticle}

\usepackage{amssymb}

\journal{Physica A}

\usepackage{hyperref}
\hypersetup{
	colorlinks=true,       
	linkcolor=blue,          
	citecolor=blue,        
	filecolor=blue,      
	urlcolor=blue           
}

\newcommand{\GR}{G_{\mathrm{R}}}
\newcommand{\Grr}{G_{\mathrm{rr}}}
\newcommand{\Gpr}{G_{\mathrm{pr}}}
\newcommand{\Gqr}{G_{\mathrm{qr}}}
\newcommand{\Gqrd}{G_{\mathrm{qrd}}}
\newcommand{\Gqrnd}{G_{\mathrm{qrnd}}}

\newcommand{\Wrr}{W_{\mathrm{rr}}}
\newcommand{\Wpr}{W_{\mathrm{pr}}}
\newcommand{\Wqr}{W_{\mathrm{qr}}}
\newcommand{\Wqrnd}{W_{\mathrm{qrnd}}}

\begin{document}

\begin{frontmatter}



\title{Interdependence of sectors of economic activities for world countries from the reduced Google matrix analysis of WTO data}


\author[utinam]{C\'elestin Coquid\'e}
\author[utinam]{Jos\'e Lages}
\author[lpt]{Dima L.Shepelyansky}

\address[utinam]{Institut UTINAM, OSU THETA, Universit\'e de Bourgogne Franche-Comt\'e, CNRS, Besan\c con, France}
\address[lpt]{Laboratoire de Physique Th\'eorique, IRSAMC, Universit\'e de Toulouse, CNRS, UPS, 31062 Toulouse, France}

\begin{abstract}
We apply the recently developed reduced Google matrix algorithm for the
analysis of the OECD-WTO world network of economic activities.
This approach allows to determine interdependences and interactions
of economy sectors of several countries, including China, Russia and USA,
properly taking into account the influence of all other world countries and 
their economic activities.
Within this analysis we also obtain the sensitivity of
economy sectors and EU countries to petroleum activity sector.
We show that this approach takes into 
account multiplicity of network links 
with economy interactions
between countries and activity sectors 
thus providing more rich information compared to the usual 
export-import analysis. 
\end{abstract}




\end{frontmatter}

\%

\section{Introduction}
\label{sec:1}

The statistical data of UN COMTRADE \cite{comtrade}
and the World Trade Organization (WTO) Statistical Review 2018 \cite{wto2018}
demonstrate all the complexity of international trade and economic relations
between world countries. The world economy
and trade are mutually interacting that makes the analysis
of their development very important but also complicated  \cite{krugman2011}. 
Thus developed advanced mathematical tools are required for 
scientific analysis of this complex system.
Usually in economy research this analysis 
uses the matrix tools of Input-Out transactions  broadly
aplied in economy starting from the fundamental works of Leontief 
\cite{leontief1,leontief2}
with their more recent developments described in \cite{miller2009}.

An additional line of research is related with the  complex networks
emerged in the last two decades with the
development of modern society generating enormous communication and social networks including the World Wide
Web (WWW), Wikipedia, Facebook, Twitter 
(see e.g. \cite{dorogovtsev}). Here, the
PageRank algorithm, developed by Brin and Page in 1998
\cite{brin} for the WWW information retrieval, became at the
mathematical foundation of the Google search engine (see
e.g. \cite{meyer}). The algorithm constructs the Google matrix $G$
of Markov chain transitions between network nodes  and
allows to rank billions of web pages of the WWW. The efficient applications
of Google matrix analysis to various directed networks
have been demonstrated in \cite{rmp2015}.

The application of Google matrix approach to the World Trade Network (WTN) 
was pushed forward in  \cite{wtn1,wtn2} using the UN COMTRADE database \cite{comtrade}
for almost 50 years of world trade. In addition to the PageRank algorithm
it was shown that the analysis of the network  with the 
CheiRank algorithm \cite{linux,wikizzs} for inverted flow links 
plays for trade equally important role. Indeed, the  PageRank 
probabilities of nodes are on average proportional to the number of
ingoing links characterizing the trade import while the CheiRank probabilities 
are on average  proportional to the number of
outgoing links thus characterizing trade export \cite{wtn1,wtn2}. 
Since both export and import are of principal importance for
trade this clearly shows the importance of combined PageRank and CheiRank analysis.
The new element of Google matrix approach is a democratic 
treatment of world countries independently of their
richness being different from the usual Import and Export
ranking. At the same time the contributions of various
products are taken to be proportional to their trade
volume contribution in the exchange flows.

While the UN COMTRADE database contains an enormous 
amount of information for all UN countries and thousands of 
products (commodities) it does not contain data
on an effective transformation of one product
to another one during various economic activities.
The Google matrix analysis of the World Network of Economic Activities (WNEA) constructed from the OECD--WTO trade in
value-added database has been reported in \cite{escaith}.
In a certain sense activities (or sectors) are similar to 
products in the WTN. However, for
the WTN of UN COMTRADE there is exchange between countries but there
is no exchange between industries and commodities.
In contrast to that in real economy certain products 
are transferred to each other (e.g.
metal and plastic are used for production of cars).
Thus the approach developed in the OECD--WTO WNEA (2013) incorporates the
transitions between activity sectors thus representing the
economic reality of world activities in a more correct 
manner as discussed in \cite{escaith}.

Hence, the new important element of the WNEA is the presence of
direct interactions
between sectors of economic activity.
Thus it is interesting to know what are the differences and
similarities of transformations between sectors 
of specific countries taking into account their exchange 
with other world countries. To obtain the interactions between sectors
of a given country one should take into account their direct interactions (links)
but also all indirect pathways of product transformations 
due to their exchange flow via the rest of the world.
The most appropriate mathematical tool 
for the extraction of this direct and indirect interactions
is the reduced Google matrix algorithm (REGOMAX)
invented in \cite{greduced}. The efficiency of this
approach has been demonstrated for various Wikipedia networks
including interactions between politicians \cite{politwiki},
world universities \cite{wrwu2017},
largest world banks \cite{wikibank}
and biological networks of protein-protein interactions \cite{proteinplos}.
Recently the REGOMAX analysis of UN COMTRADE data allowed to obtain
the influence of petroleum and gas on EU countries \cite{wtn3}.
Thus here we use the REGOMAX approach to obtain the interdependence
of sectors of economic activities for world countries 
from the OECD-WTO TiVA database (WTO data) studied already
in \cite{escaith}.

We note that there is a number of 
investigations of the world trade network data sets 
(see e.g. 
\cite{vespignani,fagiolo1,hedeem,fagiolo2,garlaschelli2010,benedictis}).
The main different feature of our approach is the use of Google matrix methods
which characterize both import and export flows taking into account the whole
transfer chain including the importance of nodes.
The analysis of these both import and export directions is rather rare,
e.g., in \cite{plosjapan} the hubs and authorities of the WTN has been
studied but we think that the Google matrix analysis with
PageRank, CheiRank and REGOMAX tools  characterizes the economy activities in
on a more deep and detailed level. The matrix analysis of financial risk
and economy complexity already demonstrated its efficiency
for undirected flows \cite{bouchaud,guhr,marsili}. 
However, the financial and trade flows are directional 
and thus we hope that the Google matrix
tools used here will find further useful applications
for financial flows and economy complexity.
The recent studies of directed interbank interactions
\cite{goetz,meller} indicate on possible interesting applications
of Google matrix analysis to financial bank flows.

We suppose that the REGOMAX algorithm, developed from 
the physical problems of scattering \cite{greduced}, 
can become a useful tool
for research in the field of econophysics \cite{econophysics}.

\section{Methods and data description}
\label{sec:2}

\begin{table}[!ht]
	\centering
	\resizebox{0.8\columnwidth}{!}{
		\begin{tabular}{|c|c|l|} 
			\hline 
			& OECD ICIO Category & ISIC Rev. 3 correspondence \\ 
			\hline 
			\hline 
			
			1 &  C01T05 AGR	& \shortstack[l]{ 01 - Agriculture, hunting and related service activities \\
				02 - Forestry, logging and related service activities \\
				05 - Fishing, operation of fish hatcheries and fish farms; service activities incidental to fishing} \\ \hline
			2 &  C10T14 MIN	& \shortstack[l]{ 10 - Mining of coal and lignite; extraction of peat \\
				11 - Extraction of crude petroleum and natural gas; service activities incidental to oil and gas extraction excluding surveying \\
				12 - Mining of uranium and thorium ores \\
				13 - Mining of metal ores \\
				14 - Other mining and quarrying} \\ \hline
			3 &  C15T16 FOD & \shortstack[l]{ 15 - Manufacture of food products and beverages \\
				16 - Manufacture of tobacco products} \\ \hline
			4 &  C17T19 TEX	& \shortstack[l]{ 17 - Manufacture of textiles \\
				18 - Manufacture of wearing apparel; dressing and dyeing of fur \\
				19 - Tanning and dressing of leather; manufacture of luggage, handbags, saddlery, harness and footwear} \\ \hline
			5 &  C20 WOD	& \shortstack[l]{ 20 - Manufacture of wood and of products of wood and cork, except furniture; \\
				Manufacture of articles of straw and plaiting materials} \\ \hline
			6 &  C21T22 PAP	& \shortstack[l]{ 21 - Manufacture of paper and paper products \\
				22 - Publishing, printing and reproduction of recorded media} \\ \hline
			7 &  C23 PET	& 23 - Manufacture of coke, refined petroleum products and nuclear fuel \\ \hline
			8 &  C24 CHM	& 24 - Manufacture of chemicals and chemical products \\ \hline
			9 &  C25 RBP	& 25 - Manufacture of rubber and plastics products \\ \hline
			10 & C26 NMM	& 26 - Manufacture of other non-metallic mineral products \\ \hline
			11 & C27 MET	& 27 - Manufacture of basic metals \\ \hline
			12 & C28 FBM	& 28 - Manufacture of fabricated metal products, except machinery and equipment \\ \hline
			13 & C29 MEQ	& 29 - Manufacture of machinery and equipment n.e.c. \\ \hline
			14 & C30 ITQ	& 30 - Manufacture of office, accounting and computing machinery \\ \hline
			15 & C31 ELQ	& 31 - Manufacture of electrical machinery and apparatus n.e.c. \\ \hline
			16 & C32 CMQ	& 32 - Manufacture of radio, television and communication equipment and apparatus \\ \hline
			17 & C33 SCQ	& 33 - Manufacture of medical, precision and optical instruments, watches and clocks \\ \hline
			18 & C34 MTR	& 34 - Manufacture of motor vehicles, trailers and semi-trailers \\ \hline
			19 & C35 TRQ	& 35 - Manufacture of other transport equipment \\ \hline
			20 & C36T37 OTM	& \shortstack[l]{ 36 - Manufacture of furniture; manufacturing n.e.c. \\
				37 - Recycling} \\ \hline
			21 & C40T41 EGW	& \shortstack[l]{ 40 - Electricity, gas, steam and hot water supply \\
				41 - Collection, purification and distribution of water} \\ \hline
			22 & C45 CON	& 45 - Construction \\ \hline
			23 & C50T52 WRT	& \shortstack[l]{ 50 - Sale, maintenance and repair of motor vehicles and motorcycles; retail sale of automotive fuel \\
				51 - Wholesale trade and commission trade, except of motor vehicles and motorcycles \\
				52 - Retail trade, except of motor vehicles and motorcycles; repair of personal and household goods} \\ \hline
			24 & C55 HTR	& 55 - Hotels and restaurants \\ \hline
			25 & C60T63 TRN	& \shortstack[l]{ 60 - Land transport; transport via pipelines \\
				61 - Water transport \\
				62 - Air transport \\
				63 - Supporting and auxiliary transport activities; activities of travel agencies} \\ \hline
			26 & C64 PTL	& 64 - Post and telecommunications \\ \hline
			27 & C65T67 FIN	& \shortstack[l]{ 65 - Financial intermediation, except insurance and pension funding \\
				66 - Insurance and pension funding, except compulsory social security \\
				67 - Activities auxiliary to financial intermediation} \\ \hline
			28 & C70 REA	& 70 - Real estate activities \\ \hline
			29 & C71 RMQ	& 71 - Renting of machinery and equipment without operator and of personal and household goods \\ \hline
			30 & C72 ITS	& 72 - Computer and related activities \\ \hline
			31 & C73 RDS	& 73 - Research and development \\ \hline
			32 & C74 BZS	& 74 - Other business activities \\ \hline
			33 & C75 GOV	& 75 - Public administration and defense; compulsory social security \\ \hline
			34 & C80 EDU	& 80 - Education \\ \hline
			35 & C85 HTH	& 85 - Health and social work \\ \hline
			36 & C90T93 OTS	& \shortstack[l]{ 90 - Sewage and refuse disposal, sanitation and similar activities \\
				91 - Activities of membership organizations n.e.c. \\
				92 - Recreational, cultural and sporting activities \\
				93 - Other service activities} \\ \hline
			37 & C95 PVH	& 95 - Private households with employed persons \\ \hline
		\end{tabular}
	} 
	\caption{List of sectors considered by Input/Output matrices from WTO-OECD database, their correspondence to the ISIC UN classification is also given.}
	\label{tab1}
\end{table}

\subsection{WNEA data sets}

As in \cite{escaith} we use the data available from the OECD-WTO
TiVA database released in May 2013 which covers years
1995, 2000, 2005, 2008, 2009. The network contains
$N_c = 58$ world countries (57 plus 1 for the Rest Of the World ROW)
given in Table 1 in \cite{escaith}.
It contains the main world countries.
We do not reproduce this list here since we concentrate
our analysis only on several leading countries
with the main emphasis on USA, Russia and China.
We use for countries ISO-3166-1 alpha-3 code available at Wikipedia.
There are also $N_s = 37$ sectors of economic activities given in 
Table~\ref{tab1}.
The sectors are classified according to the International
Standard Industrial Classification of All Economic Activities (ISIC) Rev.3 described at \cite{comtrade} and Wikipedia. 
We take into account all 37
sectors, noting that the sectors $s = 1,\dots,21$
represent production activities while $s = 22,\dots, 37$ represent service activities.
We concentrate our analysis on sectors $s=1,\dots,21$.
The total size of the Google matrix is $N=N_c N_s = 58 \times 37 = 2146$.
The main analysis is presented for year 2008.
Additional data for other years are available at
\cite{ourwebpage}.

\subsection{Google matrix construction for WNEA}

In the following we use the approach developed in \cite{wtn2,escaith}
to construct the Google matrix of financial transfers 
between economic activity sectors
of different countries. We keep the notations used in \cite{escaith}. 

From the WTO data we construct the matrix $M_{cc^\prime,ss^\prime}$
of money transfer between nodes expressed in USD of current year
\begin{equation}
M_{cc^\prime,ss^\prime} = \mbox{transfer
	from country $c^\prime$, sector $s^\prime$ to  country $c$, sector $s$}.
\label{eq1}
\end{equation}
Here the country indexes are $c,c^\prime\in\{1,\ldots,N_c\}$ and 
activity sector indexes are $s,s^\prime\in\{1,\ldots,N_s\}$
with $N_c=58$ and $N_s=37$. 
The whole matrix size is $N=N_c \times N_s = 2146$.
Here each node represents a pair of
country and activity sector, a link gives a transfer from
a sector of one country to another sector of another country.
We construct the matrix $M_{cc^\prime,ss^\prime}$
from the TiVA Input/Output tables using the transposed
representation so that the volume of products or sectors flows 
in a column from line to line; for a given country $c$ we exclude possible exchanges $(c,s)\rightarrow(c,s)$ from a sector $s$ to itself. The matrix construction of $M_{cc',ss'}$ highlights the trade exchange flows intra- and inter-countries.

The Google matrices $G$ and $G^*$ are real
matrices with non-negative elements of size $N\times N$ defined as
\begin{equation}
G_{ij}= \alpha S_{ij}+(1-\alpha) v_i ,\;\mbox{ and, }\;
{G^*}_{ij}=\alpha {S^*}_{ij}+(1-\alpha) v^*_i,
\label{eq2}
\end{equation}
where $N=N_c\times N_s$, $\alpha$ is the damping factor ($0<\alpha<1$), 
and $\mathbf{v}$ is a 
positive column vector called a \emph{personalization vector} 
with the normalization $\sum_i v_i=1$ \cite{meyer,wtn2}.
We note that the usual Google matrix corresponds to 
a personalization vector with $v_i=1/N$. 
Here as in \cite{wtn1,wtn2}, we fix $\alpha=0.5$
noting that a variation of $\alpha$ in a range $0.5$--$0.9$
does not significantly affect the probability distributions of PageRank 
and CheiRank vectors \cite{meyer,rmp2015,wtn1}. 
The  personalization vector is taken from the vector representing
the exchange weight of each sector as it is described in \cite{escaith}
(for the multiproduct WTN the same choice of this vector is 
described in \cite{wtn2,wtn3}). 
As in \cite{wtn2,escaith} we call this approach
the Google Personalized Vector Method (GPVM).

The matrices $S$ and $S^*$  are built from money matrices ${M}_{cc^\prime,ss^\prime}$ as
\begin{eqnarray}
\nonumber
S_{i,i^\prime}&=&\left\{\begin{array}{cl}   
{M}_{cc^\prime,ss^\prime}/V^*_{c^\prime s^\prime}& 
\mbox{    if } V^*_{c^\prime s^\prime}\ne0\\ 
1/N & \mbox{    if } V^*_{c^\prime s^\prime}=0\\ 
\end{array}\right.\\
S^*_{i,i^\prime}&=&\left\{\begin{array}{cl}   
M_{{c^\prime}c,s^\prime s}/V_{c^\prime s^\prime}& 
\mbox{    if } V_{c^\prime s^\prime}\ne0\\ 
1/N & \mbox{    if } V_{c^\prime s^\prime}=0\\ 
\end{array}\right.
\label{eq3}
\end{eqnarray}
where
$i^{(\prime)}=s^{(\prime)}+(c^{(\prime)}-1)N_s\in\{1,\dots,N\}$. 
We have also defined $V_{cs}=\sum_{c's'} M_{cc',ss'}$  and $V^*_{cs}=\sum_{c's'} M_{c'c,s's}$ which are the total volume of import and export for the sector $s$ of country $c$.
The sum of elements of each column of 
$S$ and $S^*$ is normalized to unity and  $G$, $G^*$, $S$, and $S^*$
belong to the class of Google matrices.
The import properties are characterized by $S$ and $G$, 
and export properties by $S^*$ and $G^*$. 

The PageRank and CheiRank are
right eigenvectors of matrices $G$ and $G^*$ with 
eigenvalue $\lambda=1$. Their components are nonzero real numbers
with their sum being normalized to unity. The components give
the probabilities to find a random seller (surfer) on a given node
after a long walk over the network. The PageRank $K$ and CheiRank $K^*$ indexes
are defined from the decreasing ordering of probabilities of PageRank vector $P$ 
and of CheiRank vector $P^*$ as
$P(K)\ge P(K+1)$ and $P^*(K^*)\ge P^*(K^*+1)$ with $K,K^*=1,\dots,N$.
Since we have countries and sectors it is convenient to
use two indexes $c,s$ probabilities $P_{cs}$ and ${P^*}_{cs}$
with $1 \leq c \leq 58$ and $1\leq s \leq 37$.
The sum over all sectors gives the probabilities 
$P_c$ and ${P^*}_c$ for each country.

\subsection{Reduced Google matrix for WNEA}

The REGOMAX algorithm, proposed in \cite{greduced}, 
is described in detail in \cite{politwiki}.
Here we give the main elements of this method
keeping the notations of \cite{politwiki,wtn3}.

The reduced Google matrix $\GR$ is constructed for a selected subset of
$N_r$ nodes. 
The construction is based on concepts of scattering theory 
used in different fields including mesoscopic physics, nuclear physics, and  
quantum chaos. It captures, in a matrix of size $N_r \times N_r$,
the full contribution of direct and indirect pathways,  happening 
in the global network of $N$ nodes,  between  $N_r$ selected nodes of interest. 
The PageRank probabilities of the $N_r$ nodes are the same 
as for the global network with $N$ nodes,
up to a constant factor taking into account that 
the sum of PageRank probabilities over $N_r$
nodes is unity. The $(i,j)$-element of $\GR$ 
can be interpreted as the probability for a random seller (surfer) starting at 
node $j$ to arrive in node $i$ using direct and indirect interactions. 
Indirect interactions refer to pathways composed in part of nodes different 
from the $N_r$ ones of interest.    
The computation steps of $\GR$ offer 
a decomposition of $\GR$ into matrices that clearly distinguish 
direct from indirect interactions: 
$\GR = \Grr + \Gpr + \Gqr$ \cite{politwiki}.
Here $\Grr$ is generated  by the direct links between selected 
$N_r$ nodes in the global $G$ matrix with $N$ nodes. The matrix 
$\Gpr$ is usually rather close to 
a matrix in which each column reproduces 
the PageRank vector $P_r$. 
Due to that $\Gpr$ does not bring much information about direct 
and indirect links between selected nodes.
The interesting role is played by $\Gqr$. It takes 
into account all indirect links between
selected nodes appearing due to multiple pathways via 
the $N$ global network nodes (see~\cite{greduced,politwiki}).
The matrix  $\Gqr = \Gqrd + \Gqrnd$ has diagonal ($\Gqrd$)
and non-diagonal ($\Gqrnd$) parts where $\Gqrnd$
describes indirect interactions between different nodes.
The explicit formulas of the mathematical and numerical computation 
methods of all three matrix components of $\GR$ are given 
in \cite{greduced,politwiki,wtn3}.

\subsection{Sensitivity of economy balance}
\label{subs:balance}

As in \cite{wtn3}, within the REGOMAX approach
we determine the whole economy balance of a given country
with PageRank and CheiRank probabilities
as $B_c = (P^*_c - P_c)/(P^*_c + P_c)$.
The sensitivity of 
economy balance $B_c$ to the price of, e.g., the sector $s$ of petroleum (sector [C23 PET])
 can be obtained 
by the change of the corresponding money volume
flow related to this sector by multiplying it by $(1+\delta)$,
computing all rank probabilities and then
the derivative $D(s \rightarrow c)=dB_c/d\delta$. 
This approach was used in \cite{wtn2,wtn3}.

We can also use the same procedure to 
determine for a given country
the sensitivity of its economy sector balance 
$B_{cs} = (P^*_{cs} - P_{cs})/(P^*_{cs} + P_{cs})$ to the 
price variation of the petroleum sector $s$. Then the sensitivity
of a sector $s'$ of a given country $c$ to another sector
$s$ is defined as $D(s \rightarrow cs') = d B_{cs'}/d \delta$.

\begin{figure}[!t]
	\centering
	\includegraphics[width=0.9\columnwidth]{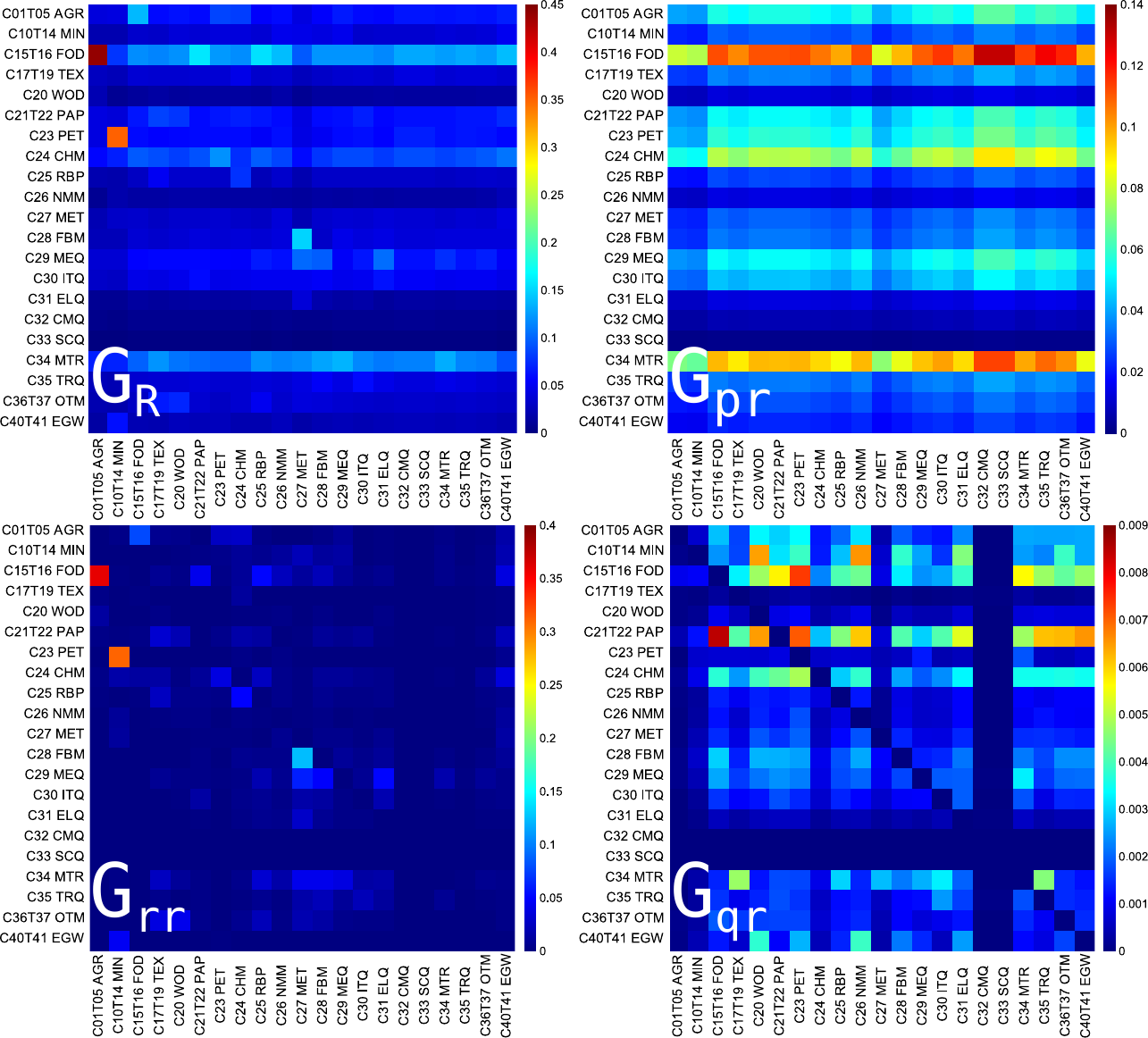}
	\caption{Density plot of reduced Google matrix 
		for import or PageRank direction:
		$\GR$ (top left), $\Gpr$ (top right), $\Grr$ (bottom left) 
		and $\Gqr$ without diagonal elements (bottom right). 
		The matrices are computed 
		for a set of reduced nodes composed of $N_r=21$ sectors ($s=1,\dots,21$) of USA for the year 2008.
		The corresponding matrix weights are: $\Wpr = 0.813817$, $\Wrr = 0.155258$, 
		$\Wqr = 0.030925$ and $\Wqrnd = 0.027383$. 
	}
	\label{fig1}
\end{figure}

\begin{figure}[!t]
	\centering
	\includegraphics[width=0.9\columnwidth]{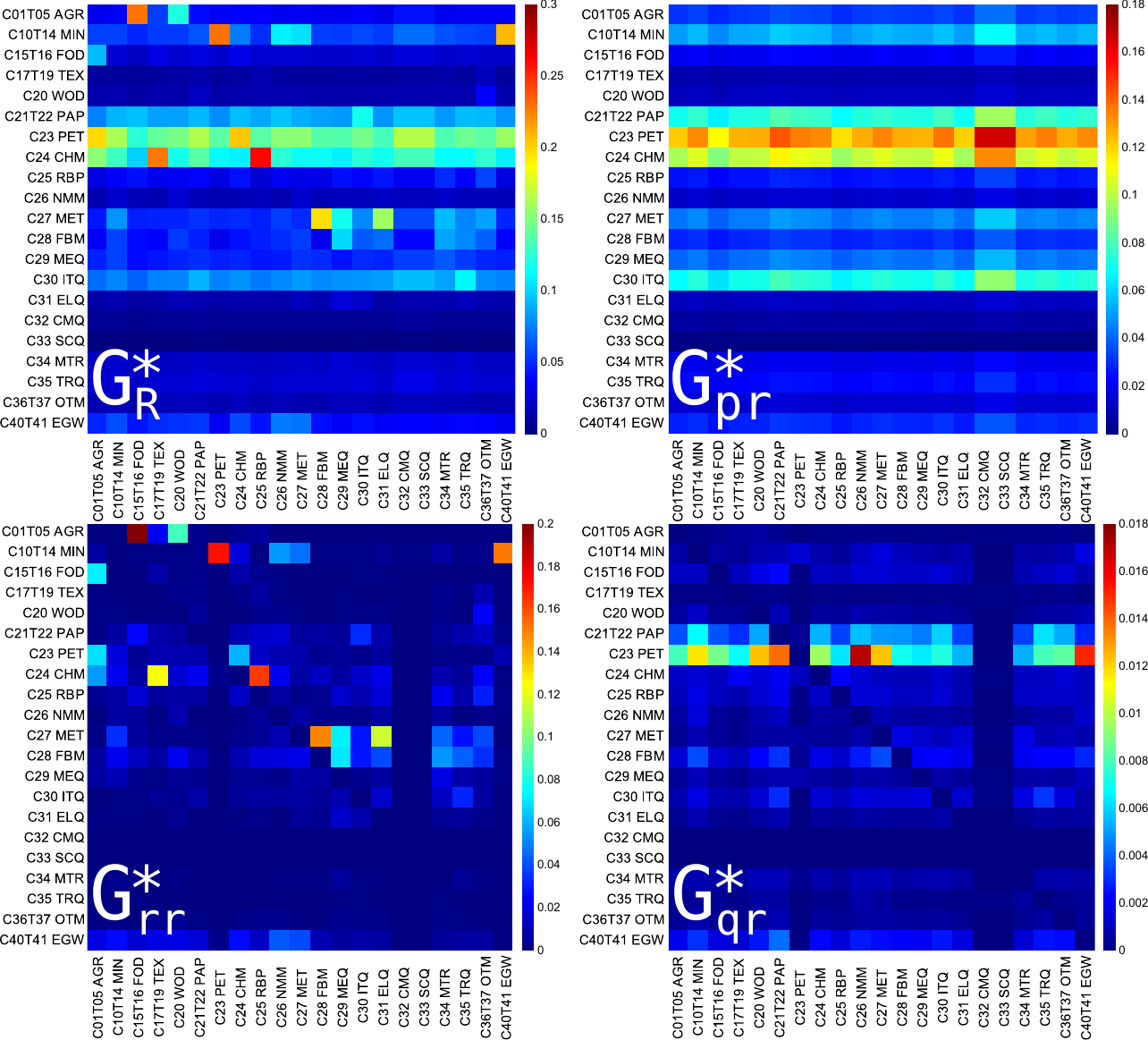}
	\caption{Density plot of reduced Google matrix
		for export or CheiRank direction:
		$\GR^*$ (top left), $\Gpr^*$ (top right), $\Grr^*$ (bottom left) 
		and $\Gqr^*$ without diagonal elements (bottom right). The matrices are computed 
		for a set of reduced nodes composed of $N_r=21$ sectors ($s=1,\dots,21$) of USA for the year 2008.
		The corresponding matrix weights are: $\Wpr^* = 0.78968$, $\Wrr^* = 0.18289$, 
		$\Wqr^* = 0.02743$ and $\Wqrnd^* = 0.02554$. 
	}
	\label{fig2}
\end{figure}

\begin{figure}[!t]
	\centering
	\includegraphics[width=0.9\columnwidth]{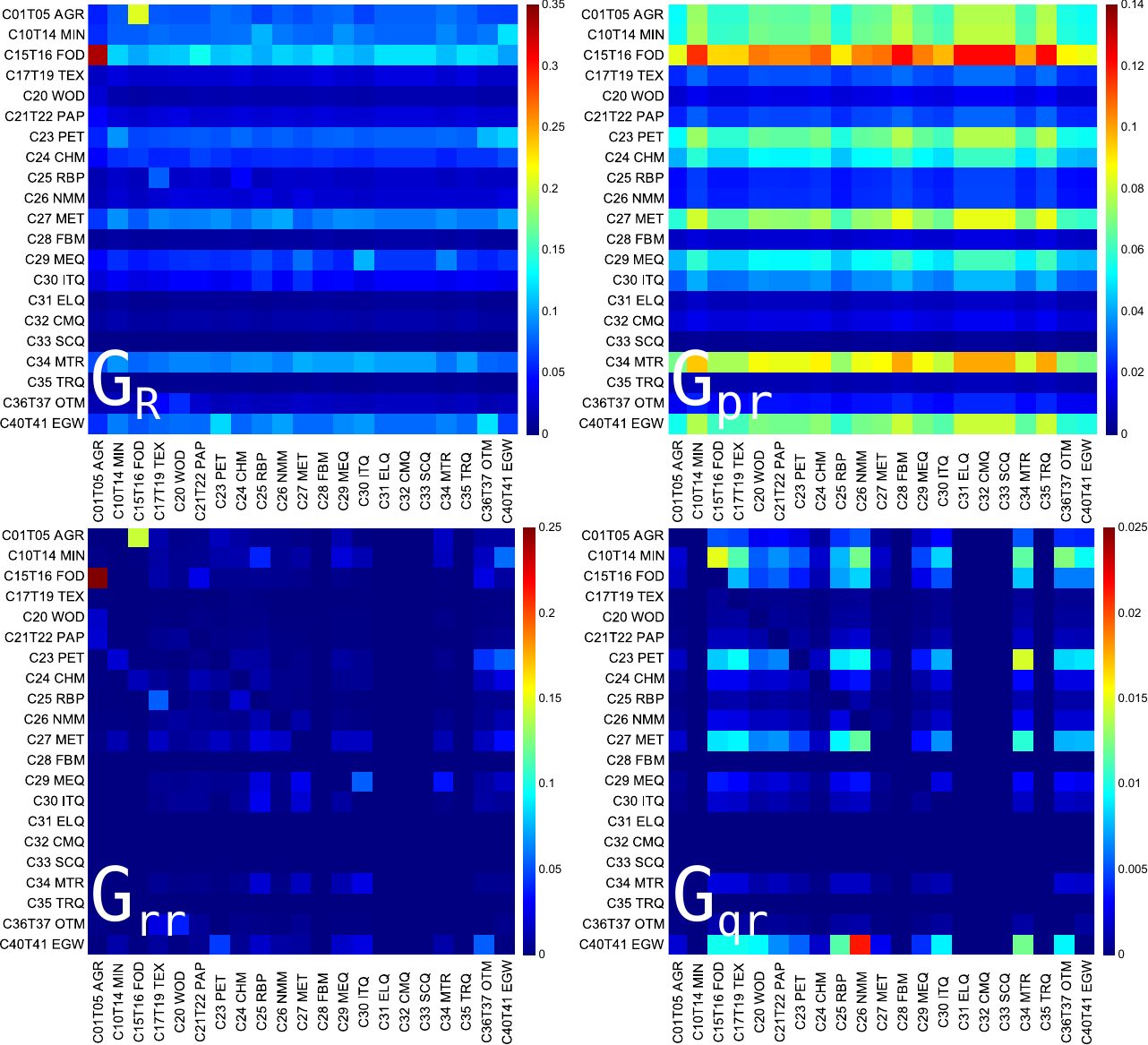}
	\caption{Same as in Fig.~\ref{fig1} for Russia RUS in 2008.
		The corresponding matrix weights are: $\Wpr = 0.851677$, $\Wrr = 0.112809$, 
		$\Wqr = 0.035514$ and $\Wqrnd = 0.033682$. 
	}
	\label{fig3}
\end{figure}

\begin{figure}[!t]
	\centering
	\includegraphics[width=0.9\columnwidth]{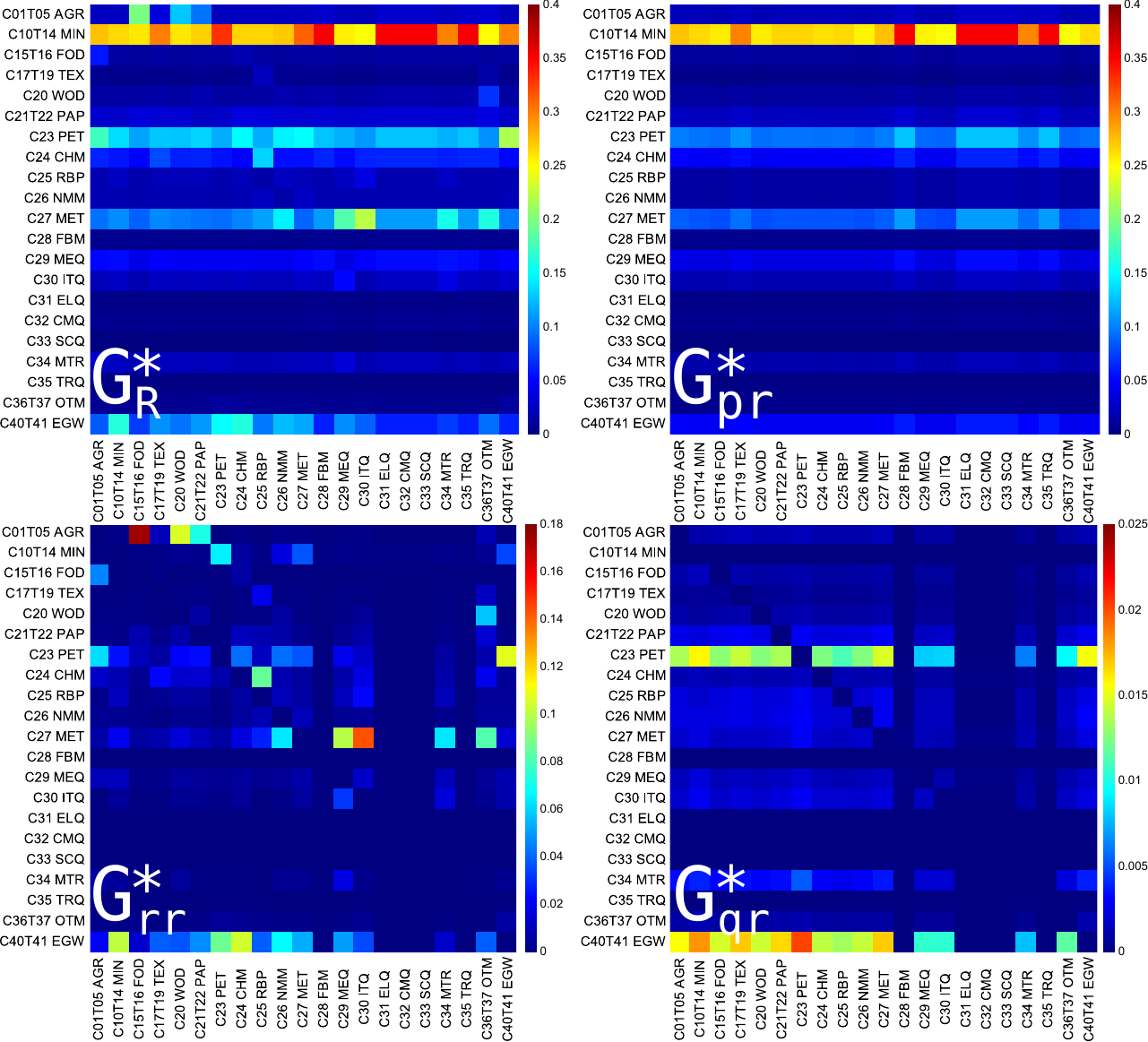}
	\caption{Same as in Fig.~\ref{fig2} for Russia RUS in 2008.
		The corresponding matrix weights are: $\Wpr^* = 0.804255$, 
		$\Wrr^* = 0.159634$, $\Wqr^* = 0.036111$ and $\Wqrnd^* = 0.033377$. 
	}
	\label{fig4}
\end{figure}

\begin{figure}[h!]
	\centering
	\includegraphics[width=0.9\columnwidth]{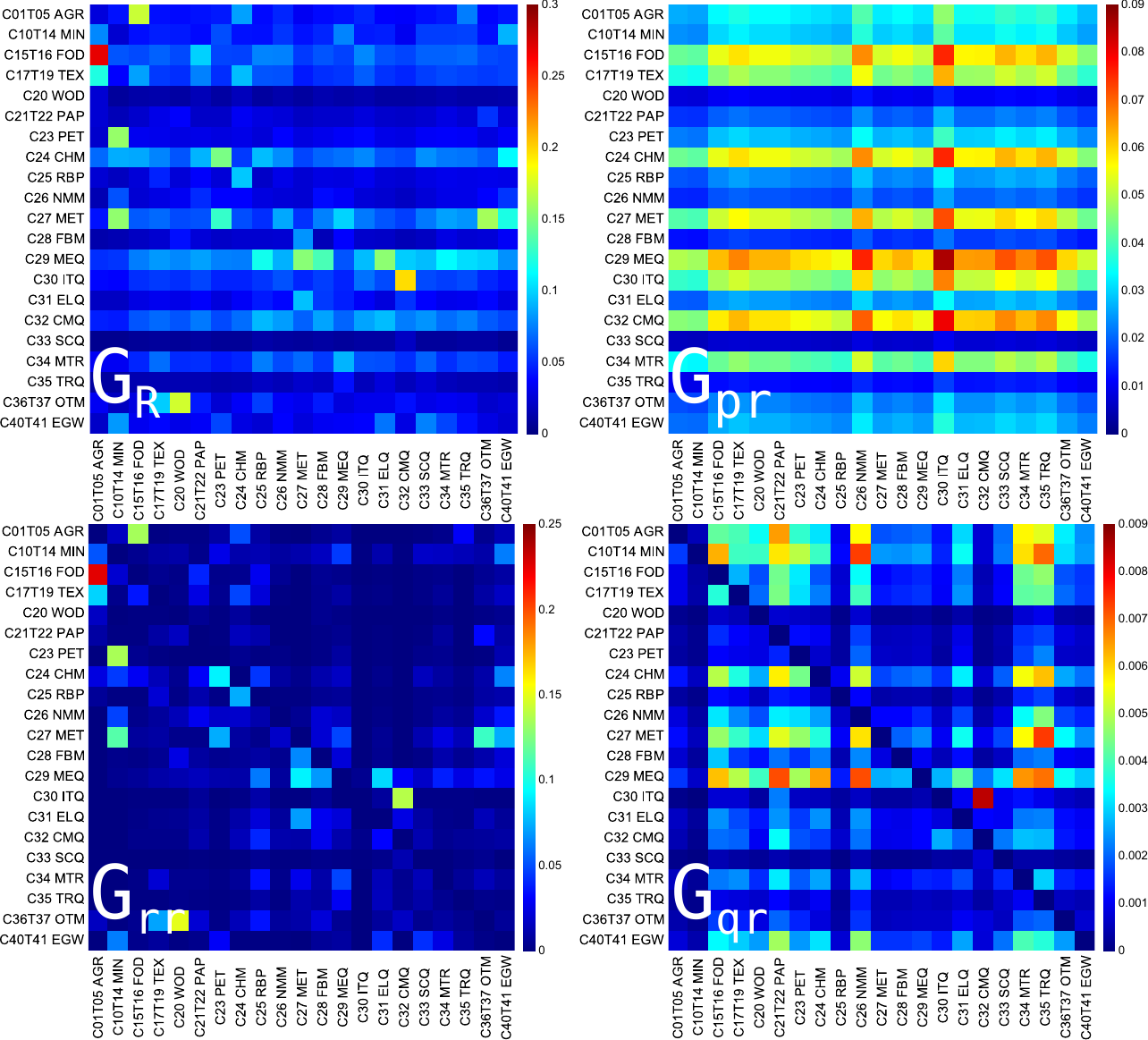}
	\caption{Same as in Fig.~\ref{fig1} for China CHN in 2008.
		The corresponding matrix weights are: $\Wpr = 0.698164$, $\Wrr = 0.263683$, 
		$\Wqr = 0.038153$ and $\Wqrnd = 0.035547$. 
	}
	\label{fig5}
\end{figure}

\begin{figure}[h!]
	\centering
	\includegraphics[width=0.9\columnwidth]{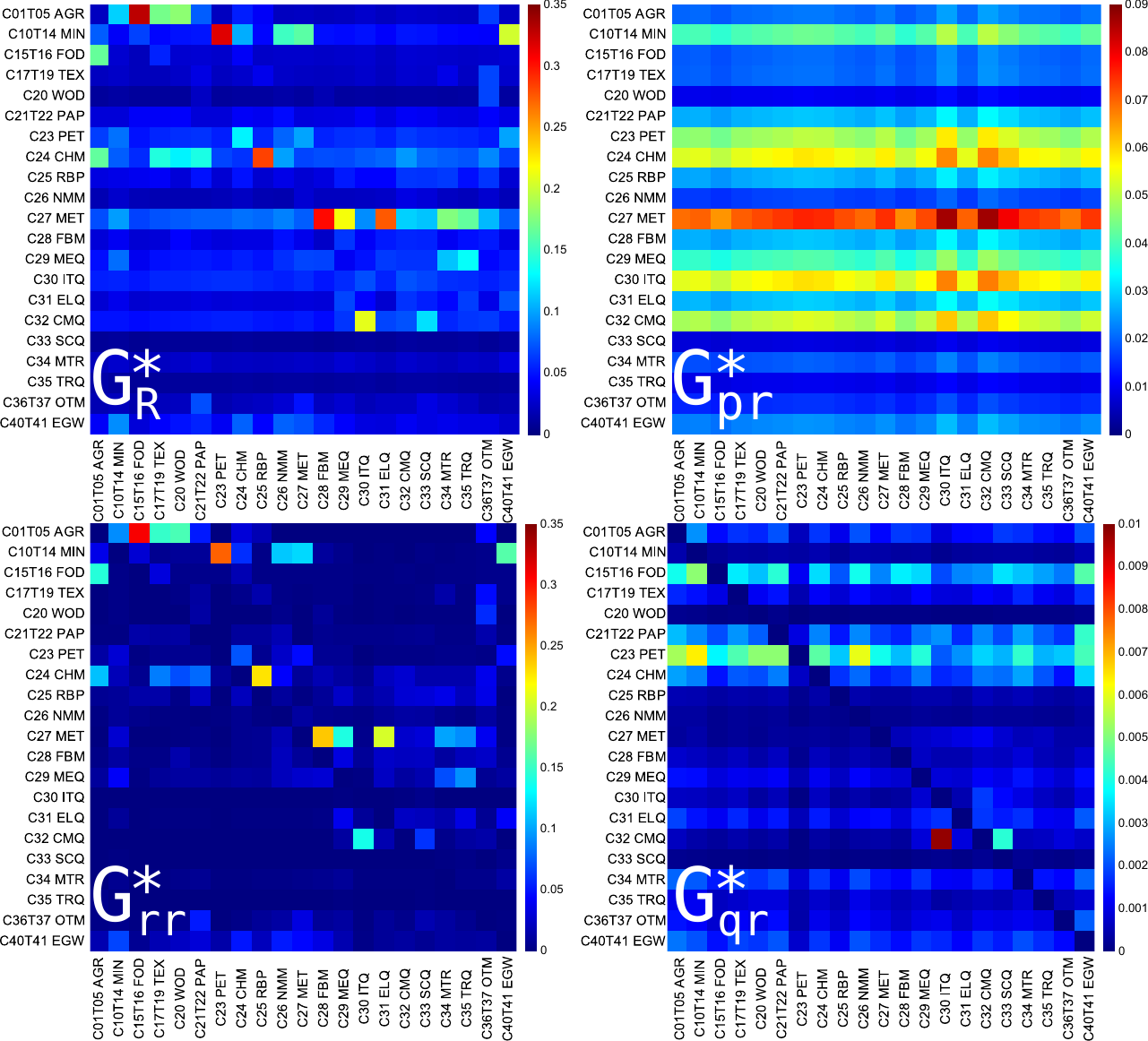}
	\caption{Same as in Fig.~\ref{fig2} for China CHN in 2008.
		The corresponding matrix weights are: $\Wpr^* = 0.647087$, $\Wrr^* = 0.326402$, 
		$\Wqr^* = 0.026511$ and $\Wqrnd^* = 0.024648$. 
	}
	\label{fig6}
\end{figure}

\section{Results}
\label{sec:3}

\subsection{Interdependence of USA economy sectors}
\label{USAsectors}

The reduced Google matrices $\GR$ and $\GR^*$ and their three matrix components 
for PageRank and CheiRank directions with $N_r=21$ sectors of economy activity 
($s=1,...,21$ in Table~\ref{tab1})
of USA are shown in Fig.~\ref{fig1} and Fig.~\ref{fig2} respectively. 
As in \cite{politwiki,wtn3} it is useful to characterize each matrix component
by their weights  $\Wpr$, $\Wrr$, 
$\Wqr$ (and $\Wqrnd$) corresponding to $\Gpr$, $\Grr$, $\Gqr$ (and $\Gqrnd$).
For each component the weight is defined as the sum of all matrix elements divided by 
the matrix size $N_r$. By definition $ \Wpr +  \Wrr + \Wqr =1$.
For USA matrix weights are given in the captions of Fig.~\ref{fig1} and Fig.~\ref{fig2}.
For Wikipedia networks (see, e.g., \cite{politwiki,wrwu2017,wikibank}) one usually has the weights $\Wpr \approx 0.95$
and $\Wrr \approx \Wqr \approx 0.025$ \cite{politwiki}. Here the situation 
is more similar to the WTN case \cite{wtn3}
and the weight of $\Wqr$ remains rather small 
while $\Wrr$ is by a factor 3-10 larger than in Wikipedia.
As for WTN  \cite{wtn3} we attribute this to a significantly larger
number of links per node in the WTN and WNEA global networks.

The most strong matrix elements in $\GR$ show the interdependence of sectors of USA
for import or PageRank direction (see Fig.~\ref{fig1}). Here we see the dominance of
interaction from $s=1$, [C01T05 AGR], related to agriculture,
to $s=3$, [C15T16 FOD], related to manufacture of food products.
Indeed, the agricultural activity produces food used by all people 
that makes this link so strong.
Another strong link is from $s=2$, [C10T14 MIN], related to mining, to 
$s=7$, [C23 PET], related to manufacture of coke, refined petroleum products.
Indeed, coke and petroleum are produced by mining 
and they play the important role in USA economy.
The third by strength link is from 
$s=11$ [C27 MET] (manufacture of basic metals) to $s=12$ [C28 FBM]
(manufacture of fabricated metal products) that is also very natural.
These three links are also well present among direct links
in $\Grr$ that corresponds to the importance of direct links in the WNEA
discussed above. Interestingly, other links, from [C21T22 PAP], paper and paper product, and from [C25 RBP], rubber and plastics products, to [C15T16 FOD], which are not so strong in direct matrix component $\Grr$ are enhanced in $\GR$, illustrating the role of indirect interactions. Indeed food products industry use products of paper and plastic industries for packaging.

In the $\Gpr$ matrix component there are 3 dominant horizontal lines
at  [C15T16 FOD], [C34 MTR], motor vehicles, and [C24 CHM], chemicals and chemical products, which are pointed by
majority of sectors. These three sectors, which are major sector activities using products of many others, are at the top 3 PageRank
positions of USA sectors.

The $\Gqr$ matrix component highlights hidden links between USA economic sectors.
Among the most pronounced hidden interactions in $\Gqr$ we note
[C15T16 FOD] pointing to [C21T22 PAP], [C23 PET] pointing to
[C15T16 FOD] and [C21T22 PAP]. It is clearly understandable that food and paper industries use indirectly petroleum products. Concerning the [C15T16 FOD] to [C21T22 PAP] link, according to $\Grr$ food industry directly points to agriculture, and of course paper industry uses sivivculture. This is one of the many possible indirect paths linking [C15T16 FOD] to [C21T22 PAP].

For the export or CheiRank direction the results are shown in Fig.~\ref{fig2}
with the strongest links from $s=9$ [C25 RBP] (manufacture of rubber and plastics products)
to $s=8$ [C24 CHM] (manufacture of chemicals),
from $s=4$ [C17T19 TEX] (manufacture of textiles) to $s=8$ [C24 CHM],
$s=7$ [C23 PET] to  $s=2$ [C10T14 MIN].
Here again the dominant contribution is given by $\Grr^*$
but the strength of final amplitudes are slightly
corrected by $\Gpr^*$ and $\Gqr^*$ contributions which mainly highlights the fact that  petroleum and chemistry industries are the main exporters to other economic activity sectors.

The amplitudes of all matrix elements of $\GR$ for USA are available at \cite{ourwebpage}.
The results for other years can be  also found there for USA, RUS, CHN.

\subsection{Interdependence of Russia economy sectors}
\label{RUSsectors}

The reduced Google matrices for Russia for PageRank (import)
and CheiRank (export) directions are shown in Fig.~\ref{fig3}
and \ref{fig4} respectively. They are constructed in the same manner
as Figs.~\ref{fig1} and \ref{fig2} for USA. 

For the reduced Google matrix $\GR$ of Russian economic sectors with PageRank (import) direction in Fig.~\ref{fig3}
the strongest link is between $s=1$ [C01T05 AGR] and $s=3$ [C15T16 FOD] similar to USA case. We note here that the inverted link is weaker but also present in $\GR$ and in $\Grr$. This is certainly due to the fact that food industry also produces products for animal used in agriculture.
In $\Gpr$ the most pronounced horizontal line is  
$s=3$ [C15T16 FOD] highlighting the fact that this industry uses indirectly products of almost all the other economic sectors; it is followed by the line of
$s=18$ [C34 MTR] (manufacture of motor vehicles)
and $s=21$ [C40T41 EGW] (electricity, gas).
Among indirect links the strongest is
from $s=10$ [C26 NMM] (manufacture of other non-metallic mineral products)
to $s=21$ [C40T41 EGW] (electricity, gas).

For the reduced Google matrix $\GR^*$ of Russian economic sectors with CheiRank (export) direction in Fig.~\ref{fig4}
there is the dominance of lines related to
$s=2$ [C10T14 MIN] (mining), followed with weaker intensities by [C23 PET] (petroleum), [C27 MET] (basic metals), and [C40T41 EGW] (electricity, gas). This picture is rather different from USA case
in Fig.~\ref{fig2}. Although there are only very few weak direct links pointing to [C10T14 MIN], the mining sector is very important, since through the network of exports it strongly acts (via almost only indirect interactions) to every sectors of the Russian economy.
For hidden links in $\Gqr^*$
the dominant line is for [C40T41 EGW] (electricity, gas).

\subsection{Interdependence of China economy sectors}
\label{CHNsectors}

Interdependence of economy sectors of China CHN 
for PageRank (import) and CheiRank (export)
directions is presented with the reduced Google matrix and its components
in Fig.~\ref{fig5} and \ref{fig6} respectively.

For the reduced Google matrix $\GR$, shown in Fig.~\ref{fig5}, 
there are strong links between $s=1$ [C01T05 AGR] and $s=3$ [C15T16 FOD] similar to 
USA and RUS cases. In addition there is a strong transition from
$s=16$ [C32 CMQ] (manufacture of communication equipment)
to $s=14$ [C30 ITQ] (manufacture of computing machinery).
This correspond to strong CHN production of TV, computers and other
communication related products. In the matrix component $\Gpr$ 
there are strong lines for [C15T16 FOD], [C24 CHM], [C29 MEQ] (machinery and equipment), and [C32 CMQ]. We note that contrarily to USA and RUS the food sector does not dominate alone as top importer, chemistry, communication, computing, and machinery industries play also an important role as they also use indirectly products of many Chinese economic sectors.
For hidden links in the matrix component $\Gqr$ the strongest matrix element points from
[C32 CMQ] to [C30 ITQ]. Many other links with a slightly weaker intensity are also highlighted by the $\Gqr$ matrix component, but these are quite weak in comparison with the highest intensities in the reduced Google matrix $\GR$.

For the reduced Google matrix $\GR^*$, shown in Fig.~\ref{fig6}, 
the strongest matrix elements are from [C15T16 FOD] to [C01T05 AGR],
[C28 FBM] (manufacture of fabricated metal products)
to [C27 MET] (manufacture of basic metals),
[C23 PET] (petroleum) to [C10T14 MIN] (mining of coal etc.),
[C25 RBP] (rubber and plastic) to [C24 CHM], and [C23 ELQ] (electrical machinery and apparatus) to [C27 MET].
For  $\Gpr^*$ the strongest horizontal line of transitions
is for [C27 MET] (manufacture of basic metals).
Among the indirect matrix elements of $\Gqr^*$
the strongest link is from  [C30 ITQ] to [C32 CMQ].

\subsection{Intersensitivity of economy sectors}

The above results show specific dependencies between
economy sectors for USA, RUS, and CHN.

Here we choose 10 countries (USA, RUS, CHN, DEU, FRA, ITA, GBR, JAP, KOR, IND)\footnote{Excepting KOR these countries are among the top 10 importers according to PageRank algorithm applied to the global Google matrix $G$; KOR is ranked at the twelfth position.} for which we determine the balance sensitivity of each economical activity sectors to a price variation of a specific sector.

For a given country, the balance sensitivity of a sector $s'$ to an infinitesimal increase of sector $s$ product prices is $D(cs \rightarrow cs') = dB_{cs'}/d \delta$. The balance of economic sector $s$ is defined as $B_{cs} = \left(P^{*}_{cs} - P_{cs}\right)/\left(P^{*}_{cs}+P_{cs}\right)$, see subsection \ref{subs:balance} for details.

\begin{figure}[t!]
	\centering
	\includegraphics[width=0.9\columnwidth]{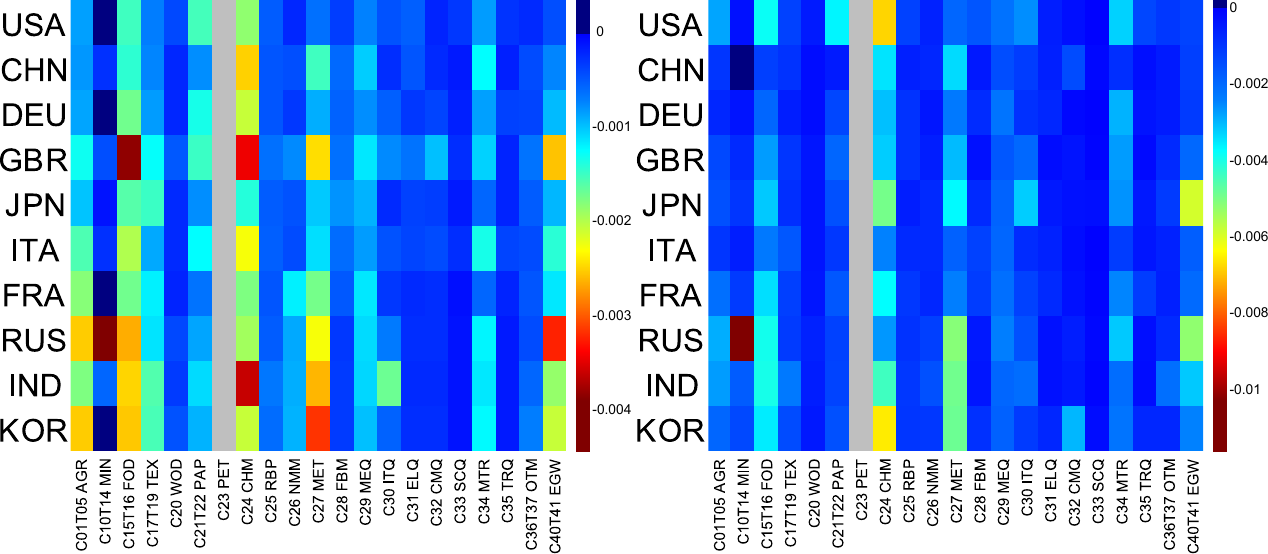}
	\caption{Sector balance sensitivity to [C23 PET] sector for year 1995 (left) 
		and 2008 (right); horizontal axis represents the sector index in data order; 
		vertical axis represents the country index in PageRank order for the given year.
		For each couple $(s,c)$ we modify the link from ([C23 PET], $c$) 
		towards $(s,c)$ and compute the $(s,c)$ balance sensitivity, $D(c\mbox{[C23 PET]},cs)$. 
                 Grey column represents self sensitivity (not shown).}
	\label{fig7}
\end{figure}

\begin{figure}[h!]
	\centering
	\includegraphics[width=0.9\columnwidth]{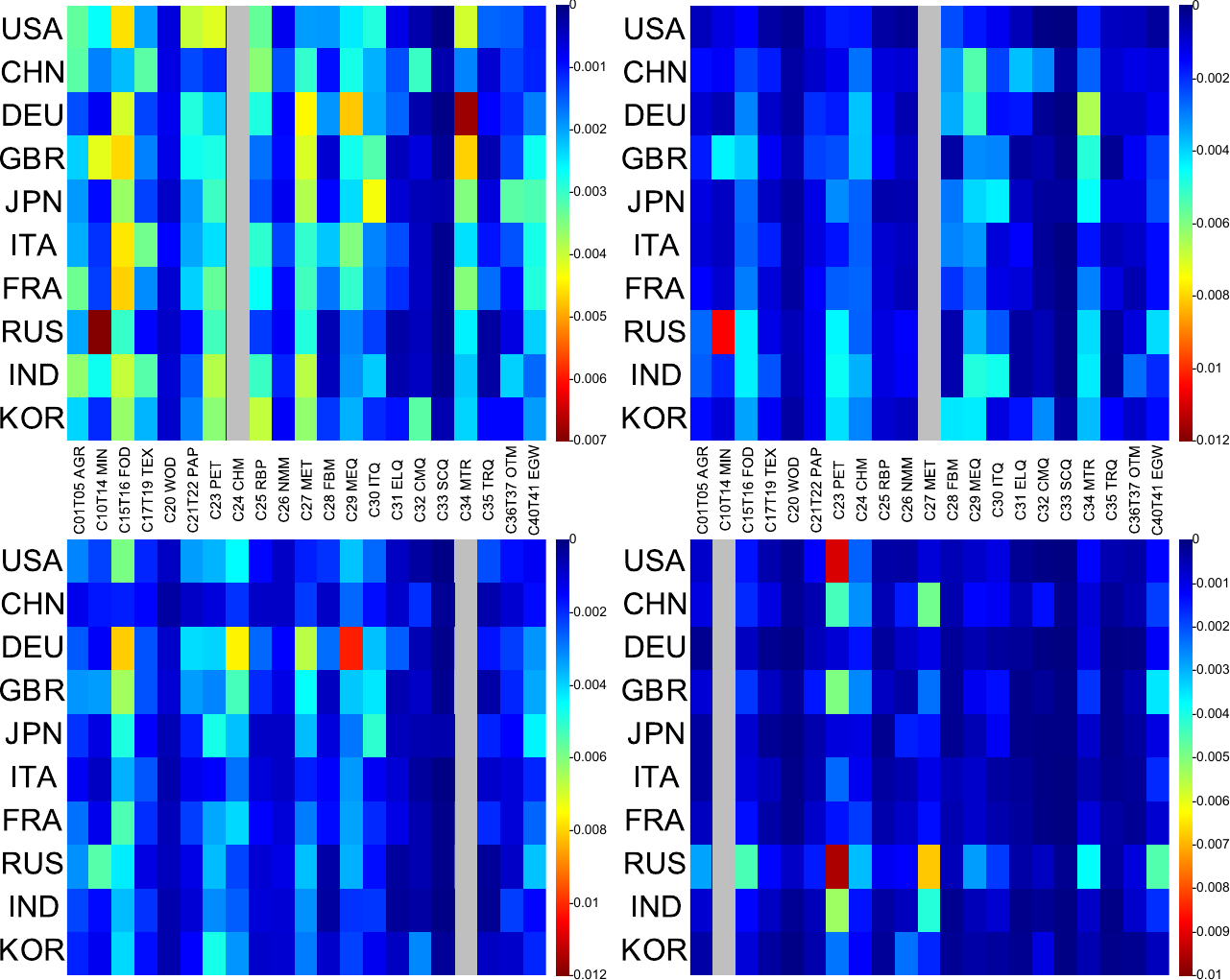}
	\caption{Sector balance sensitivity to [C24 CHM] (top left), [C27 MET] (top right), [C34 MTR] (bottom left), and [C10T14 MIN] (bottom right) sectors for year 2008; horizontal axis represents the sector index in data order; vertical axis represents the country index in PageRank order for the given year.
	For each couple $(s,c)$ we modify the link from $(s',c)$ towards $(s,c)$ and compute the $(s,c)$ balance sensitivity, $D(cs',cs)$. Grey column represents self sensitivity (not shown).}
	\label{fig8}
\end{figure}

In Fig.~\ref{fig7}, we show a map of balance sensitivity in respect to sector
[C23 PET] (petroleum products)
for years 1995 and 2008.
The maximal absolute value of the balance sensitivity $D$ is increased
approximately by a factor 3 from 1995 to 2008
showing an increased dependence of economy sectors on petroleum.
Partially it can be attributed to price growth for petroleum from 1995 to 2008
(changed by a factor $\sim3$, even $\sim5$ taking April 2008 as reference).
For both years and for any of the considered countries, we observe that some economic sectors, 
such as [C31 ELQ] (electrical machinery and apparatus), [C33 SCQ] (medical, precision and 
optical instruments, watches and clocks), [C35 TRQ] (transport equipment), and [C20 WOD], 
are almost insensitive to the petroleum products sector. Inversely, 
the most sensitive economic sectors to [C23 PET] sector are [C24 CHM] (chemicals), [C27 MET] (basic metals), [C40T41 EGW] (electricity, gas, hot water supply), [C15T16 FOD] (food), and mostly in 1995 [C01T05 AGR] (agriculture). Indeed activities of these industries directly use petroleum products. For every countries, the sector [C10T14 MIN] (mining) is robust from 1995 to 2008 except for Russian mining sector the sensitivity of which goes from -0.0045 in 1995 to -0.012 in 2008. This is a peculiarity of Russian mining sector which appears strongly dependent of Russian petroleum sector.
Same data as in Fig.~\ref{fig7} are represented in
Fig.~\ref{fig7b} but with the same color scale for 1995 and 2008. In Fig.~\ref{fig7b} we globally observe that from 1995 to 2008 the [C24 CHM] sector has increased its sensitivity to [C23 PET] sector, by, e.g., a factor $\sim3$ for USA and KOR. A weaker increase of sensitivity to [C23 PET] sector can be observed for [C27 MET] and [C15T16 FOD] sectors. We also observe that Russian and Japanese [C40T41 EGW] (electricity, gas, hot water supply) sectors became from 1995 to 2008 more sensitive to their national [C23 PET] sector. From Fig.~\ref{fig7b} we can add that all economic sectors of DEU and ITA and to a somewhat less extend of FRA, GBR and CHN remain from 1995 to 2008 insensitive to their [C23 PET] economic sector.

2008 sector balance sensitivities to [C24 CHM], [C27 MET], [C34 MTR], and [C10T14 MIN] sectors are shown in Fig.~\ref{fig8}. Among these economic sectors, [C24 CHM] sector have the broadest impact to other economic sectors. The strongest sensitivities to 
[C24 CHM] and [C27 MET] sectors concern Russian [C10T14 MIN] and German [C34 MTR] sectors. German economy is the most affected by the [C34 MTR] sector, particularly [C29 MEQ], [C15T16 FOD], and [C24 CHM] sectors. The most sensitive economic sectors to [C10T14 MIN] sector are [C23 PET] and [C27 MET] sectors, particularly Russian [C23 PET] and [C27 MET] sectors and US [C23 PET] sector.

\begin{figure}[!t]
	\centering
	\includegraphics[width=0.9\columnwidth]{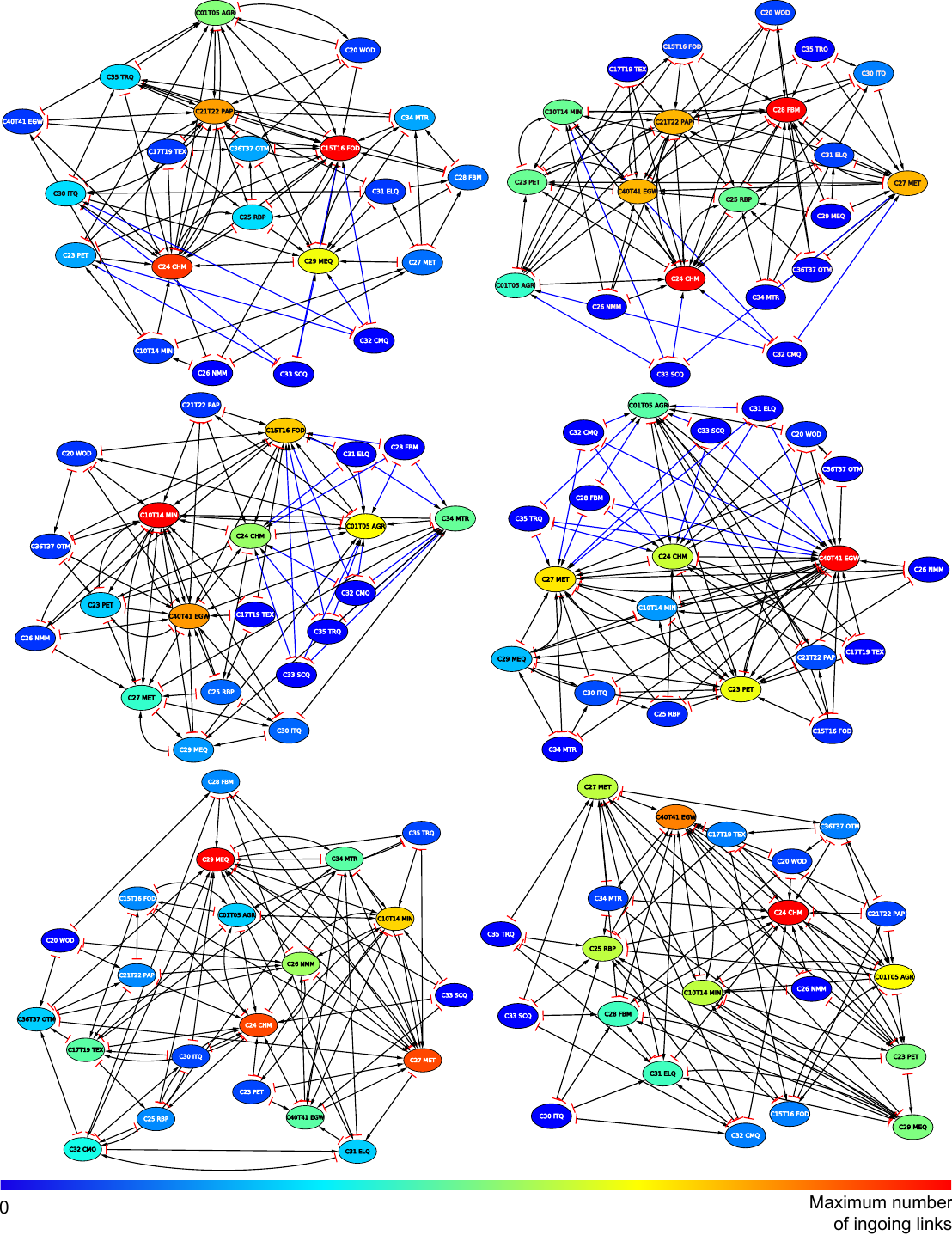}
	\caption{Reduced networks of economic sectors of USA (top row), RUS (central row), and CHN (bottom row) obtained from the corresponding \textit{import} reduced Google matrices $\GR$ (left panel) and \textit{export} reduced Google matrices $\GR^*$ (right panel) for year 2009.
	For each country, the reduced networks have been computed for a set of 21 major economic sectors. From each of them we draw the 4 strongest outgoing links. Node labels are sector codes from Tab. \ref{tab1}. The color of a node corresponds to its number of ingoing links from 0 (blue color) to the maximum (red color).
	We distinguish by blue color hidden links from direct links present in 
	the raw data. Red bars represent source-side of the links and arrows represent target-side of the links.
	The networks have been plotted with radial plot algorithm in Cytoscape software \cite{cytoscape} with manual layout optimization.}
	\label{fig9}
\end{figure}

\subsection{Reduced network of economic sectors}

We construct the reduced networks of 21 economic sectors for different countries. For that purpose we use the \textit{import} reduced Google matrices $\GR$ and \textit{export} reduced Google matrix $\GR^*$ corresponding to USA, RUS, and CHN economic sectors for 2019. Examples of such reduced Google matrices are presented in Sections \ref{USAsectors}, \ref{RUSsectors} and \ref{CHNsectors} for 2008.

For a given country $c$ and for the economic sector $s$ we select the 4 links $\{(c,s)\rightarrow (c,s_i)\}_{i=\alpha,\beta,\gamma,\delta}$ giving the strongest entries in the composite matrix $\Grr+\Gqrnd$ (or $\Grr^*+\Gqrnd^*$) extracted from the reduced Google matrix $\GR$ (or $\GR^*$) associated to the 21 economic sectors $\{(c,s_i)\}_{i=1,\dots,24}$ (see Figs.~\ref{fig1}, \ref{fig2}, \ref{fig3}, \ref{fig4}, \ref{fig5} and \ref{fig6} to have an idea of the composite matrices, $\Grr+\Gqrnd$ and $\Grr^*+\Gqrnd^*$, for USA, RUS, and CHN). The networks constructed from components of the reduced Google matrix $\GR$ ($\GR^*$) will highlight import (export) capabilities of the economic sectors. Let us remind that the compact picture given by the reduced Google matrices at the level of a country comprises in fact the global information encoded in the global Google matrix of all the transactions from any sector $s$ of a country $c$ to any sector $s'$ of a country $c'$.

In Fig.~\ref{fig9} (top row) we present the reduced network of US economic sectors for import (left panel) and export (right panel) exchanges. From import point of view, we observe that the [C24 CHM] sector uses products from the broadest variety of US economic sectors as it has the maximum of ingoing links (13 out of 21 economic sector are pointing to the [C24 CHM] sector). Other economic sectors using many US resources are [C15T16 FOD] (10 out of 21), [C21T22 PAP] (10 out of 21), [C29 MEQ] (9 out of 21), and [C01T05 AGR] (7 out of 21) sectors. From export point of view, we observe that the major suppliers of the US economic sectors are (by number of ingoing links) [C26 CHM] (13 out of 21 economic sectors are supplied by [C26 CHM] sector), [C28 FBM] (13 out of 21), [C40T41 EGW] (10 out of 21), and [C27 MET] (9 out of 21). The [C24 CHM] sector seem to play an important role since it is an economic hub using products of many other economic sectors and being a supplier of many other economic sectors. From both the import and export pictures we observe that manufacture of equipment sectors, [C32 CMQ] (radio, television and communication equipment) and [C33 SCQ] (medical, precision and optical instruments, watches and clocks) are linked to other economic sectors only by hidden links, i.e., in WNEA there is no direct money exchange between these sectors and the others.

In Fig.~\ref{fig9} (middle row) we present the reduced network of Russian economic sectors for import (left panel) and export (right panel) exchanges. Here the major importers are the followinf economic sectors [C10T14 MIN] (18 out of 21), [C40T41 EGW] (12 out of 21), [C15T16 FOD] (11 out of 21). We note that the [C10T14 MIN] sector uses product of almost all the 21 considered sectors.
From the export point of view, the major exporters are the sectors of [C40T41 EGW] (21 of 21), [C27 MET] (13 out of 21), [C23 PET] (13 out of 21), [C24 CHM] (12 of 21), and [C01T05 AGR] (9 out of 21). We note that [C40T41 EGW] sector which exploit products of all the other economic sectors is very central in Russian economy since it constitutes the major economic hub.
From both import and export pictures, as in US economy, the [C33 SCQ] and [C32 CMQ] sectors are linked to other by hidden links. In addtion to these sectors, also the [C31 ELQ] (electrical machinery), [C28 FBM] (fabricated metal products), [C35 TRQ] (transport equipment) sectors intervene through hidden links.

In Fig.~\ref{fig9} (bottom row) we present the reduced network of Chinese economic sectors for import (left panel) and export (right panel) exchanges. The major exporter are the sectors of [C29 MEQ] (11 out of 21), [C24 CHM] (10 out of 21), [C27 MET] (10 out of 21),and [C10T14 MIN] (8 out of 21). The major exporters are sector of [C24 CHM] (12 out of 21), and [C40T41 EGW] (10 out of 21). As in US economy, [C24 CHM] sector is an economic hub playing a central role in Chinese economy.

\begin{figure}[!t]
	\centering
	\includegraphics[width=0.9\columnwidth]{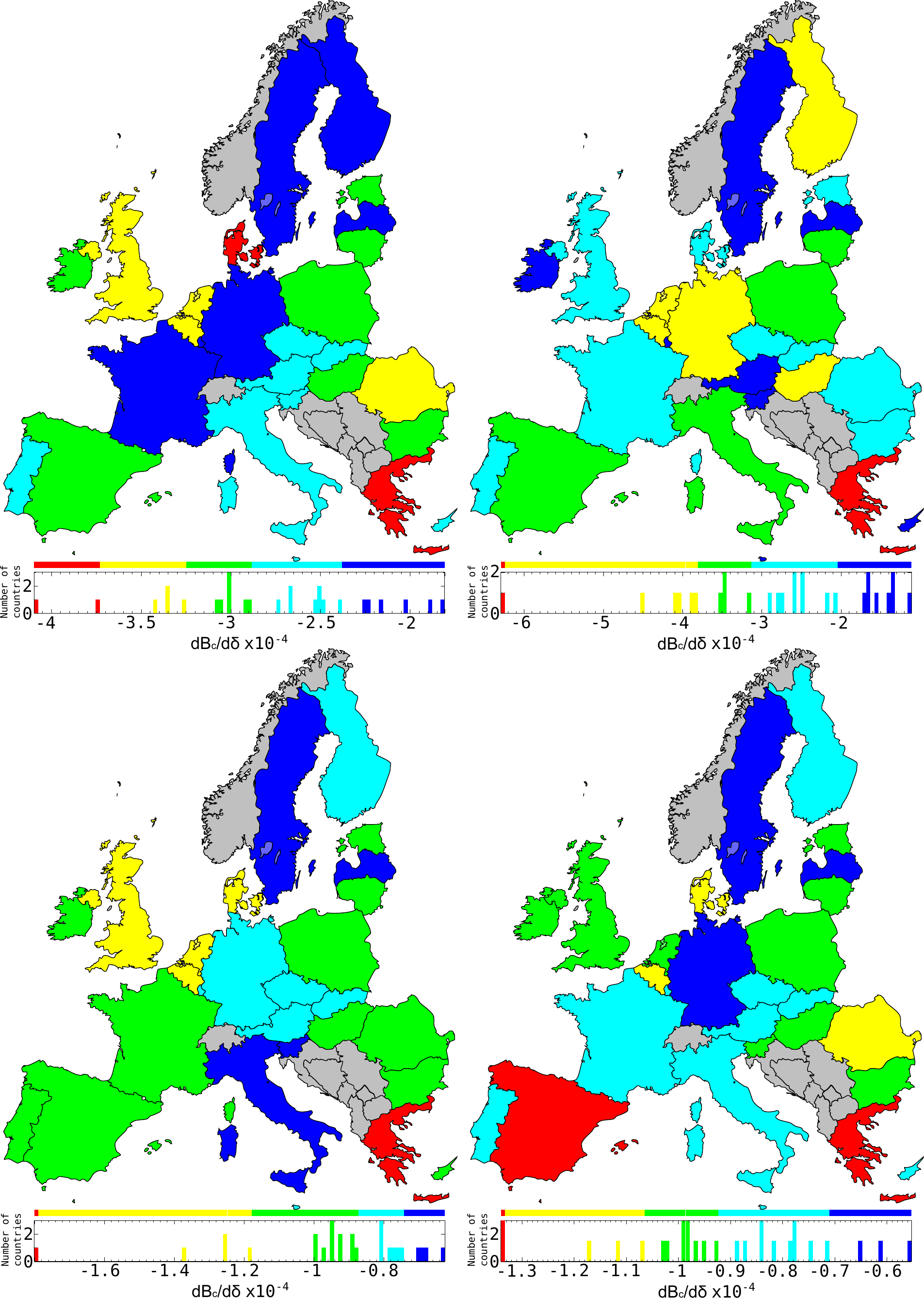}
	\caption{Balance sensitivity of the 27 EU countries in 2008 to export variation of the [C23 PET] sector of USA (top left), RUS (top right), NOR (bottom left) and SAU (bottom right). Color categories are obtained using the Jenks natural breaks classification method \cite{jenkswiki}.}
	\label{fig10}
\end{figure}

\subsection{Sensitivity of EU countries to petroleum products price}

The WNEA data combined with the REGOMAX approach allow to
study the sensitivity of country balance to a specific
economy sector. In the recent studies 
of the WTN from EU COMTRADE database \cite{wtn3}
such a sensitivity has been determined
for 27 EU countries (EU members in 2013) in respect to petroleum price.
Here for comparison we show the balance sensitivity of the same
27 EU countries in respect to sector price variation of [C23 PET] sector
related to petroleum. The results are presented in Fig.~\ref{fig10}
for [C23 PET] sectors of USA, RUS, Norway (NOR) and Saudi Arabia (SAU) in 2008.
We see that the most sensitive countries to US, Russian, and Norwegian petroleum is Greece
while to Saudian petroleum they are Greece and Spain.
Globally, the influence of USA and RUS are comparable
while the influences of NOR and SAU are by a factor 2-3 smaller.

We note that EU color maps of balance sensitivities to petroleum products from USA, RUS, and SAU are somewhat different from the WTN case shown
in \cite{wtn3} (Fig.~6 middle row panels).
We attribute this difference to the fact that the [C23 PET] sector contains
different petroleum related ISIC products while for WTN only petroleum product is used. Also WNEA comprises real inter-sector and inter-country economic exchanges. We nevertheless note that the petroleum sensitivity of Netherlands is in any case moderate to strong as in the WTN study \cite{wtn3}.

In Fig.~\ref{fig10} we observe that Sweden, Finland, and Latvia are the less sensitive to petroleum products of any of the considered suppliers. In addition to these countries we see that the less sensitive to petroleum products from US are France and Germany, from Russia are Austria, Slovenia and Ireland, from Norway is Italy, and from Saudi Arabia is Germany. In addition to Greece which is the most sensitive country to petroleum products for any of the considered suppliers, the most sensitive are Denmark to US petroleum products, and Spain to Saudian petroleum products.

\section{Discussion}

In this work we apply the reduced Google matrix (REGOMAX) 
analysis to the WNEA data and determine the interdependence 
of economy activity sectors for several countries
with the main accent to USA, Russia, and China.
There are similarities
and significant differences for
interactions of sectors of selected countries.
All three countries have strong interdependence
between [C01T05 AGR]  (agriculture) and  [C15T16 FOD] (food products) sectors
that is rather natural since all people need agriculture development
for food productions, and also between [C10T14 MIN] (mining) and [C23 PET] (petroleum products) sectors that is also very natural. For US economy we note that there are also strong interdependence between [C27 MET] (basic metals) and [C28 FBM] (fabricated metal products) sectors, and between [C25 RBP] (rubber and plastics products), [C17T19 TEX] (manufacture of textiles), and [C24 CHM] (chemistry) sectors. For Chinese economy we observe additional interdependence between [C32 CMQ] (communication equipment) and [C30 ITQ] (computing machinery), and [C23 ELQ] (electrical machinery and apparatus) to [C27 MET].
From the constructed reduced networks of economic sectors for each considered economy, we have determined an economic hub which uses a broad variety of products from the other economic sectors and supplies also many of them. For US economy [C24 CHM] sector is clearly an economic hub. For Russian economy, the [C40T41 EGW] (electricity, gas, hot water supply) sector is central as it is the main exporters to all the other sectors and in return this sector also consumes many ressources from the other sectors. For Chinese economy, as for US, [C24 CHM] sector is an economic hub.
We also determine the sensitivity of sectors of a given country to variation of price costs in a specific sector. Globally, for any of the top importer countries according to PageRank algorithm applied to WNEA (USA, RUS, CHN, DEU, FRA, ITA, GBR, JAP, KOR, IND), we observe a strong sensitivity of [C24 CHM], [C27 MET], [C40T41 EGW], and [C15T16 FOD] sectors to price increase of [C23 PET] sector products. Contrarily, [C23 ELQ], [C32 CMQ], [C35 TRQ] (transport equipment), and [C20 WOD] (wood products) sectors are the most insensitive to [C23 PET] sector. We show also the sensitivities of economic sectors to [C24 CHM], [C27 MET], [C34 MTR], and [C10T14 MIN] sectors and we determine the color map of balance sensitivities of EU countries to a price increase of products from US, Russian, Norwegian, and Saudian [C23 PET] sectors.

Our studies demonstrate that the REGOMAX method allows to 
find  inter-dependencies between economy sectors for selected countries.
The WNEA data of OECD-WTO contains transformations of production of one sector to another that is absent for multiproduct trade data of COMTRADE. Thus it would be very desirable to extend OECD-WTO data for more sectors and more recent years.
We hope that this will happen in future years.

\section*{Acknowledgments}
We thank Hubert Escaith for useful discussions.
This work was supported in part by the Pogramme Investissements
d'Avenir ANR-11-IDEX-0002-02, reference ANR-10-LABX-0037-NEXT 
(project THETRACOM);
it was granted access to the HPC resources of 
CALMIP (Tou\-lou\-se) under the allocation 2018-P0110.
This work was also supported in part by the Programme Investissements
d'Avenir ANR-15-IDEX-0003, ISITE-BFC (GNETWORKS project) and 
by Bourgogne Franche-Comt\'e region (APEX project).

%



\newpage
\appendix

\section{Additional figures}

\setcounter{figure}{0}

\begin{figure}[!ht]
	\centering
	\includegraphics[width=0.9\columnwidth]{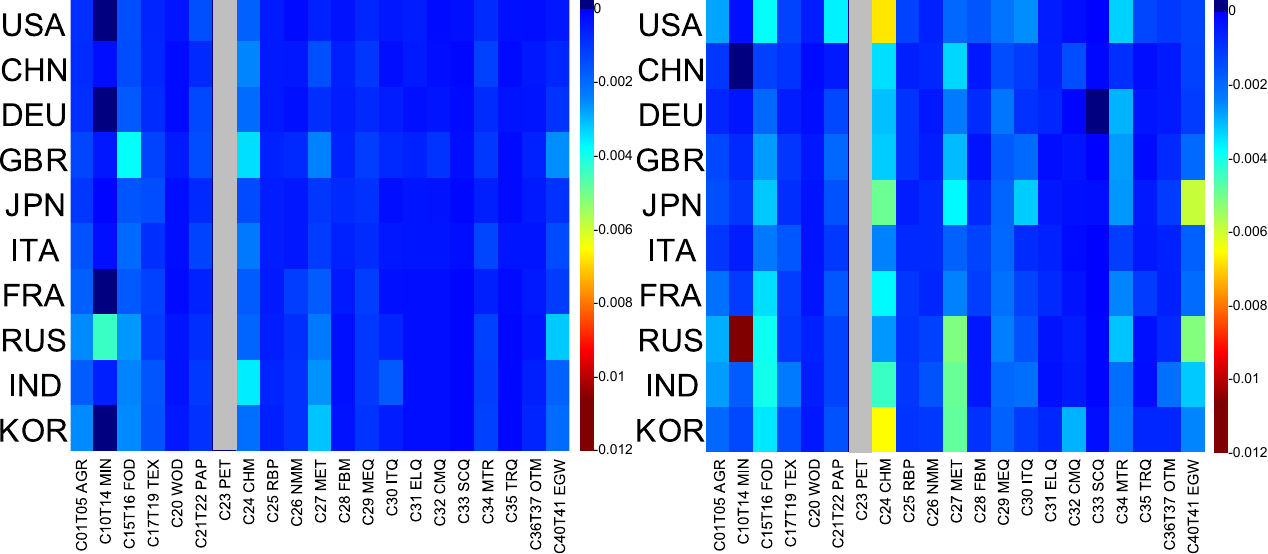}
	\caption{Same as Fig.~\ref{fig7} but with same color scale for both panels. Sector balance sensitivity to [C23 PET] sector for year 1995 (left) 
		and 2008 (right); horizontal axis represents the sector index in data order; 
		vertical axis represents the country index in PageRank order for the given year.
		For each couple $(s,c)$ we modify the link from ([C23 PET], $c$) 
		towards $(s,c)$ and compute the $(s,c)$ balance sensitivity, $D(c\mbox{[C23 PET]},cs)$. Grey column represents self sensitivity (not shown).}
	\label{fig7b}
\end{figure}

\end{document}